# Transverse shift of a beam with orbital angular momentum under reflection from a dielectric film

**N. D. Kundikova**[1,2], **K. A. Zaitsev**[1]
*1. Institute of Electrophysics, Ural Division of Russian Academy of Sciences, 106 Amundsen str., Yekaterinburg, 620016 Russia*
*2. South Ural State University, 76 Lenin Ave., Chelyabinsk, 454080 Russia*
e-mail: kundikovand@susu.ac.ru

**Abstract**
We present the results of numerical analysis of Gauss-Bessel beam reflection from a dielectric film under different angles of incidence. A transverse shift of the beam under the orbital momentum sign change is observed. The value of the shift is independent of the polarization state of the incident beam.

The investigation of reflection and refraction of laser beams at the boundary between two media show a deviation from the well known laws of reflection and refraction. For the first time the specific behavior of the reflected light beam was described in [1]. The longitudinal shift of the linearly polarized light was demonstrated under the total internal reflection. The shift value was on the order of magnitude of the light wavelength, and depended on the light polarization. This shift was experimentally observed under the linearly polarized light propagation through a planar waveguide [2, 3].

If light was circularly polarized, then the transverse shift of the center of gravity of the light beam was observed under total internal reflection, the shift value was comparable to the light wavelength and the direction of the shift depended on the sign of the circular polarization [4, 5]. Transverse shift was experimentally observed under propagation of a beam with non-uniform polarization through a multifaceted prism, namely, one half of the beam was right circularly polarized, and the second one was left circularly polarized [6, 7, 8].

New effects can be observed under a beam reflection from a metal surface. A negative Goos-Hanchen shift [9] and a transverse shift [10] were demonstrated under the beam reflection from the metal surface.

The effects described above are called spatial longitudinal Goos-Hanchen shift and spatial transverse Imber-Fedorov shift.

It was shown that the angle of reflection did not equal the angle of incidence under the light beam reflection from the glass surface [11]. The value of the observed shift was in the order of magnitude was $\sim 10^{-4}$ rad. This shift is known as the angular Goos-Hanchen shift.

Angular Imber-Fedorov shift was predicted and experimentally observed when light was reflected from the metal surface [12]. The spatial beam field distribution influences the effects that occur when the light beam reflects from the interface between two media. A new kind of transverse spatial and angular shift of the beam center of gravity was predicted for partially reflected or refracted beams at the same condition [12]. The predicted shifts can be observed only for beams with a non-zero orbital angular momentum and do not depend on the incident beam state of polarization.

Transverse splitting of the linearly polarized Gaussian beam into two beams with orthogonal circular polarizations was observed under the reflection from a thin metallic film [13]. The Gaussian beam deformation and longitudinal shift under the beam reflection from the dielectric film was considered in [14].

There are no publications devoted to investigation of reflection of light beams with orbital momentum from a thin film.

In this paper, we are reporting the transverse shift of a beam with an orbital momentum reflected from a dielectric film under the sign of the orbital momentum change. We present the results of numerical analysis of Gauss-Bessel beam reflection from a film under different angles of incidence.

Let us consider the reflection of Gauss-Bessel beams from a thin sapphire film deposited on silicon. The Gauss-Bessel beam propagation was described in the frame of the wave equation. The wave equation was solved by the spectral method, based on the two-dimensional Fourier transform. The light reflection was taken into consideration in the following way. Let $E_{GB}^{i}$ is the field of the incident beam. This field can be represented as a superposition of plane waves propagating in different directions using Fourier transform. In order to obtain the reflected field we should use the reflection coefficient dependence on an incident angle for a thin film deposited on the substrate [15]. The field of the reflected Gauss-Bessel beam $E_{GB}^r$ can be obtained using inverse Fourier transform.

Let the system under investigation has the following parameters. The beam incident from air was reflected from a sapphire film. The film was deposited on a silicon base plate. The film thickness was changed from 10 to 20 wavelength. The wavelength is equal to 0.63 mkm. The silicon permittivity is equal to 13.2. The sapphire permittivity is equal to 3.2. The angle of incidence $\alpha$ can be changed from 0 till 45 degrees.

Figure 1 shows the deformation of the incident beam intensity after reflection from the film under investigation.

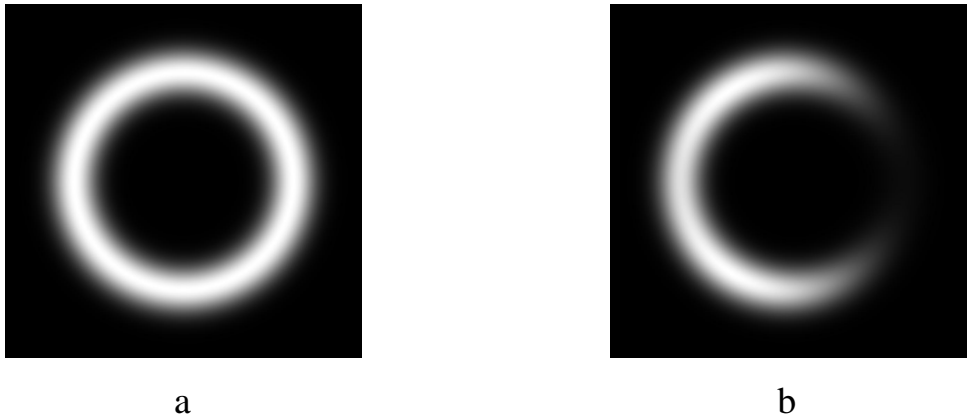

a          b

Fig.1 The deformation of the incident beam intensity after reflection from the sapphire film deposited on a silicon base plate. The dimension of a frame is 400x400 mkm. The film thickness is 12.3 mkm. The angle of the incidence $\alpha = 15°$. The topological charge $l = +1$, the beam is left circular polarized. The beam intensity distribution before (a) and after (b) reflection.

It can be easily seen from Fig.1 that the reflection results in the beam deformation in the direction of the incidence plane and in the longitudinal shift of the beam center of gravity. The value of the longitudinal sift $d_1$ is equal to -45 mkm. It is impossible to see the transverse shift of the beam center of gravity from the Fig.1 because the value of the shift is equal to 0.75 mkm. It turns out that the state of the beam polarization (spin momentum) does not influence the intensity distribution of the reflected beam, but the change of the orbital momentum sign results in the transverse shift of the beam center of gravity. The transverse shift of the beam center of gravity for the beam with $l=-1$ is equal to -0.75 mkm, so the change of the orbital momentum sign leads to shift equaled to 1.5 mkm.

The change of the angle of incidence leads to different kinds of beam intensity deformation. Figure 2 shows the deformation of the incident beam intensity after reflection for the case of $\alpha=45°$.

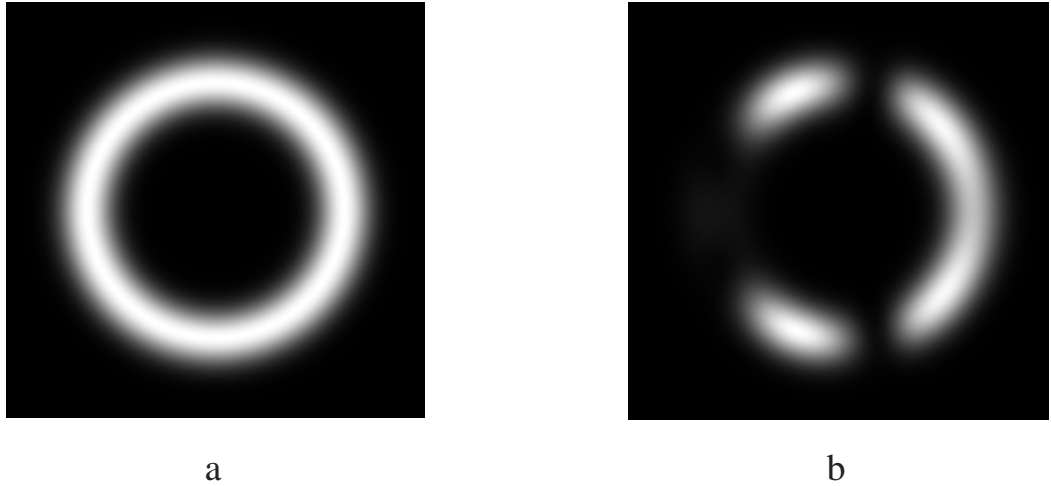

a  b

Fig.2 The deformation of the incident beam intensity after reflection from the sapphire film deposited on a silicon base plate. The dimension of a frame is 400x400 mkm. The film thickness is 12.3 mkm. The angle of the incidence $\alpha=45°$. The topological charge $l=+1$, the beam is left circular polarized. The beam intensity distribution before (a) and after (b) reflection.

It can be easily seen from Fig.2 that the change of the angle of incidence results in deformation of another kind, the beam intensity distribution is symmetrical in relation to the plane of incidence and there is the longitudinal shift of the beam center of gravity. The value of the longitudinal shift $d_1$ is equal to 34 mkm. The value of the transverse shift is equal to 0.52 mkm. As in the previous case the state of the beam polarization (spin momentum) does not influence the intensity distribution of the reflected beam and the change of the orbital momentum sign results in the transverse shift of the beam center of gravity. The transverse shift of the beam center of gravity for the beam with $l=-1$ is equal to -0.52 mkm, so the change of the orbital momentum sign leads to shift equaled to 1.04 mkm.

As a result we provide the first prediction, to the best of our knowledge, a transverse shift of a beam with orbital momentum reflected from a dielectric thin film under the change the sign of an orbital momentum. The value of the shift is independent of the polarization state of the incident beam.